\documentclass[3p]{elsarticle}

\usepackage{amsmath,amssymb}
\usepackage{epsfig,graphicx}

  \usepackage[T1]{fontenc}
  \usepackage{aecompl}

\begin{document}

\title{Observational constraints on the types of cosmic strings}

\author[rvt]{Olga S.~Sazhina\corref{cor1}\fnref{fn1}}
\ead{cosmologia@yandex.ru}
\author[focal]{Diana Scognamiglio\fnref{fn2}}
\ead{wediaaa91@hotmail.it}
\author[rvt]{Mikhail V. Sazhin\fnref{fn1}}
\ead{moimaitre@mail.ru}

\cortext[cor1]{Corresponding author}

\address[rvt]{Sternberg Astronomical Institute of Lomonosov Moscow State University, Moscow, Russia}
\address[focal]{University of Naples Federico II, Naples, Italy}

\fntext[fn1]{Russia 119992 Moscow, Universitetskiy pr. 13 (SAI MSU); tel.: +7 495 9395006; fax: +7 495 9328841
}
\fntext[fn2]{Italy 80126 Naples, via Cinthia, 6
}

\begin{abstract}
This paper is aimed at setting observational limits to the number of cosmic strings (Nambu-Goto, Abelian-Higgs, semilocal) and other topological defects (textures). Radio maps of CMB anisotropy, provided by the space mission Planck for various frequencies, were filtered and then processed by the method of convolution with modified Haar functions (MHF) to search for cosmic string candidates. This method was designed to search for solitary strings, without additional assumptions about the presence of networks of such objects. The sensitivity of the MHF method is $\delta T \approx 10 \mu K$ in a background of $\delta T \approx 100 \mu K$.  The comparison of these  with previously known results on search string network shows that strings can only be semilocal in an amount of $1 \div 5$, with the upper restriction on individual strings tension (linear density) of $G\mu/c^2 \le 7.36 \cdot 10^{-7}$. The texture model is also legal. There are no strings with $G\mu/c^2 > 7.36 \cdot 10^{-7}$.  However, comparison with the data for the search of non-Gaussian signals shows that the presence of several (up to 3) of Nambu-Goto strings is also possible. For $G\mu/c^2 \le 4.83 \cdot 10^{-7}$ the MHF method is ineffective because of unverifiable spurious string candidates. Thus the existence of strings with tensions $G\mu/c^2 \le 4.83 \cdot 10^{-7}$ is not prohibited but it is beyond the Planck data possibilities.

The same string candidates have been found in the WMAP 9-yr data. Independence of Planck and WMAP data sets serves as an additional argument to consider those string candidates as very promising. However the final proof should be given by optical deep surveys.

\end{abstract}


\maketitle

\section{Introduction}

The expansion and cooling of the Universe took place in several phases of transition, while different types of topological defects were formed. The best known topological defects are cosmic strings (CS), whose topological structure of the symmetry breaking at the phase transition ensures their stability. CS were postulated by Kibble \cite{1} and immediately became a hot issue in both theoretical physics and cosmology \cite{2} -- \cite{3}. It can be stated that topological CS are infinitely long and filamentary remnants of primordial dark energy (high energy symmetric vacuum) which formed in the early Universe and were then stretched by the cosmological expansion up to the point that, at present epoch, some CS could cross the  entire length of the observable Universe. The detection of CS would  allow us to disentangle the true underlying theoretical framework and to  extend our experimental knowledge in the GUT energy domain which is unavailable to modern and foreseen particle accelerators (i.e. $10^{14} \div 10^{16}$GeV). It would also allow us to confirm on an observational ground some of the key points of inflation theory. CS formation appears to be ubiquitous in GUT theories and in fundamental superstring theory. Therefore, the models with non-topological defects, called semilocal strings, have become quite popular (see \cite{31} and references therein). 

Active search for observational manifestations of CS, as well as probabilistic estimations of their number \cite{ve}, show that, whether they exist, CS should be very few. 

\section{Methods of CS observational search}

A CS produces the peculiar conical topology of the space-time and it may cause a detectable effects both in radio and in optical data. Modern strategies for detecting CS can be subdivided in the three main classes:
\begin{enumerate}
\item detection through gravitational lensing effects. The method is based on the extensive monitoring of deep optical surveys \cite{vil} -- \cite{6}. The aim is to find chains of pairs of virtual and particularly truncated galaxy images which a CS is expected to produce;
\item detection through the signatures left by CS on the CMB anisotropy \cite{71} -- \cite{7};
\item detection of model depended and rare features such as, for instance, the gravitational radiation from CS loops, the interactions of CS with black holes, the decay of heavy particles emitted by CS, and the interaction of two CS \cite{8}.
\end{enumerate}

We believe that the combined use of the first with second method is the most powerful strategy in CS search. Next, we will discuss only the CS search  in the Planck radio data.

\section{Search for anisotropy of CMB induced by a solitary CS by modified Haar functions}

We will discuss only relevant to astrophysical point of view long strings of cosmological size. We have tried to determine if anisotropy is generated by a solitary, long, straight CS moving with a constant velocity $v$, owning to the tension (linear density) $\mu$ against a homogeneous and isotropic background \cite{7}. 

The moving long CS may generate anisotropy in the CMB due to the simple Doppler mechanism. This is known as the Kaiser-Stebbins effect \cite{71}: a moving CS induces a relative speed between the light source and observer and causes a shift of photons.  

\begin{figure}[pH]
\centering\includegraphics[width=8.0cm, angle=90.0]{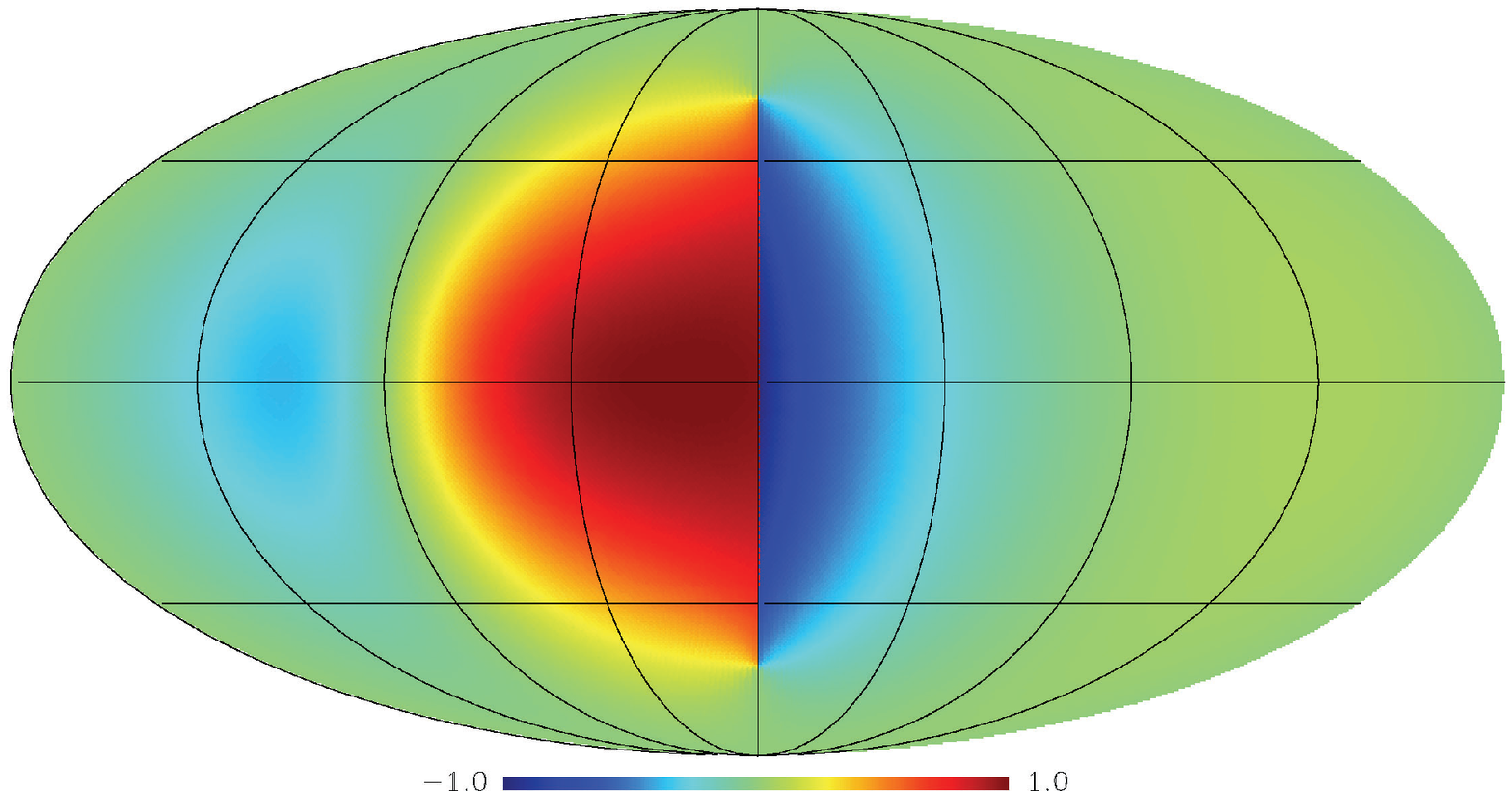}
\caption{\textit{The Molveide projection of the sky is used. It is showed simulated structure of anisotropy (temperature variation) induced by one straight CS lying on the line connecting the north and south Galactic pole. A string is moving relativistically from the left to the right. The full jump in temperature when crossing the string front is two (in dimensionless units).}}
\label{_spot}
\end{figure}

Following the standard procedure to calculate the temperature anisotropy induced by a CS (\cite{8}), we obtained:
$$
\frac{\delta T}{T} \approx 8\pi G \mu/c^2 \frac{v/c}{\sqrt{1-v^2/c^2}}.
$$
According our model (\cite{7}), an anisotropy induced by solitary CS represents a sequence of zones of decreased and increased temperature: the cold spot in front of moving string, the step-like jump, and then appearance of hot spot; finally a cold spot follows again (Fig. \ref{_spot}).

Due to the specific discontinuity structure of CS anisotropy and its low amplitude we searched its traces with a special step-like set of modified Haar functions (MHF), see Appendix I and \cite{91}. We use MHF as convolution kernel for the processing of the WMAP and Planck CMB data. The entire power of the ''signal'' (CS  anisotropy) is concentrated in a single Haar harmonic, so that an optimal marginal filter is realized.

Our method is able to detect a CS at a level of $\delta T  \approx 10 \mu K$ (\cite{91}). 

The convolution procedure is the standard technique based on the use of the most appropriate and complete orthonormal system. In our case of step-like signal the choice is MHF set.

In principle, we could expand our data in ordinary trigonometric functions of the angular coordinate of the disk (e.g., $\sin \phi$ and $\cos \phi$) and in some set of functions that are orthogonal to the disk radius (for example, Zernike polynomials, which are commonly
used when considering optical aberrations). However, generally speaking, expansion in functions (of an angular variable) of any other form implies that the signal power is distributed over all the harmonics. Thus, when expanded in trigonometric functions of
the angular anisotropy due to the CS, the amplitudes of all the harmonics will be non-zero. In other words, the signal from the CS
will be smeared out over the entire spectrum. In order for the signal to be detected, the power smeared out over all the harmonics must be ''gathered'' to make use of the full power of the signal.

For our purpose the MHF is a realization of the first harmonic of the Haar system of orthogonal functions with cyclic shift. This function is equal to $1$ in the rotation range $[0, \pi)$, and it is equal to $-1$ in the rotation range $[\pi, 2\pi)$. Since a CS could be oriented arbitrarily with respect to a grid of lines of longitude and latitude, the search for a CS at each point requires multiple convolutions with a rotation of the circle, which corresponds to a shift in the ''jump'' in the Haar function. This shift yields a new orthogonal and complete set of functions: MHF. The rotations result in a set of amplitudes. When there is a CS at a convolution point, the harmonic is maximum if a chord of the circle coincides with the position of the CS. We assigned each pixel a value equal to the corresponding maximum value of the convolution, in this way making a map of CS candidates (see Fig. (\ref{_sl})). 

Before the MHF algorithm was applied on real data we estimated its efficiency and chose the optimal convolution circle radius.

We applied MHF algorithm to process a map that was a sum of two model maps. The first map was a simulated map of the primordial CMB anisotropy that arose at the surface of last scattering. We generated 300 such maps starting from a simulated power spectrum, generated by CMBEASY \cite{q1} - \cite{q2}, a lighter and faster version of CMBFAST \cite{q3}. The second map was a pure anisotropy generated by a straight, moving CS (see Fig. (\ref{_spot}) as an example of such map). The maps of the primordial CMB anisotropy and the anisotropy generated by the moving CS were summed with a coefficient to characterize the signal-to-noise ratio. To choose an optimal circle radius for a search for CS, we performed computer simulations to obtain maps of the distribution of the harmonic amplitude for circles with various radii. We characterized the CS detection by the signal-to-noise ratio, since the CS position in the model maps was known. The amplitude at the CS location was taken to be the signal and the rms of the harmonic amplitude in the map to be the noise. Those simulations indicate that the optimal value of the convolution circle radius is from 3 deg to 5 deg. 

In order to study the efficiency of the MHF algorithm we also apply statictical methods of simulations. We created a robust set of 300 maps of sky simulating the CMB structure without any string. In Figs. (\ref{_01}) - (\ref{_03}) in Appendix II we show examples of those simulated maps, in Figs. (\ref{_1}) - (\ref{_3}) we show examples of the result of the MHF algorithm. There are less than one false string candidates in simulations (see Appendix II for details). Those statistics strongly support the efficiency of the MHF algorithm.

\section{CS candidates in Planck data} \label{cs_}

\begin{figure}[pH]
\centering\includegraphics[width=10.0cm, angle=90.0]{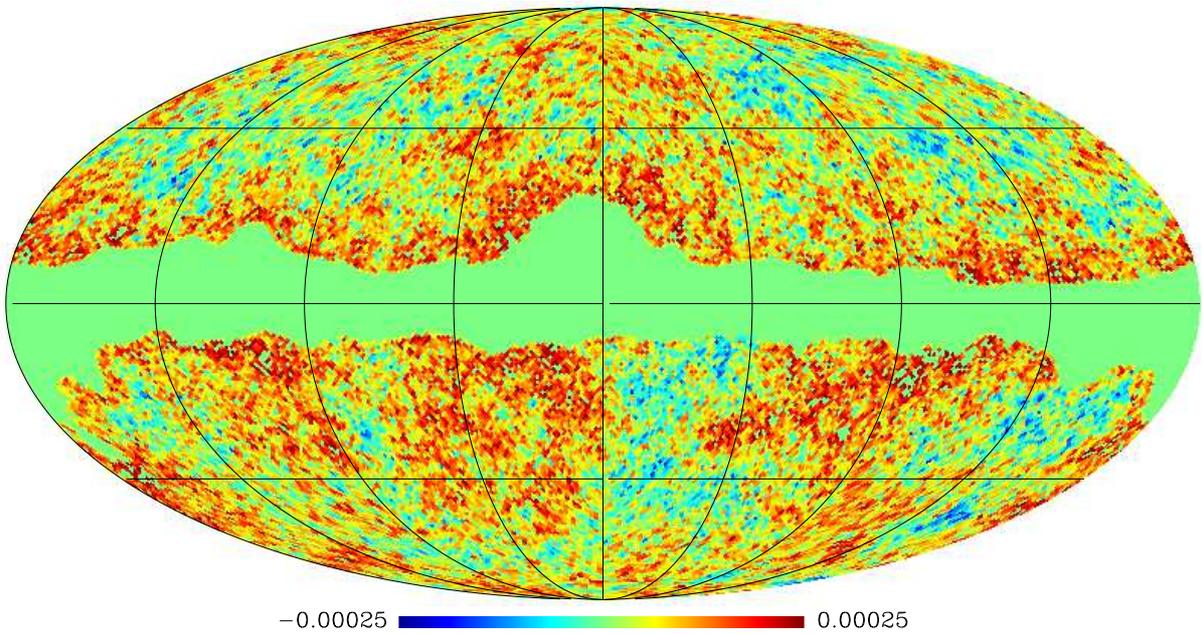}
\caption{\textit{Cleaned Planck data (143GHz, units are [K]). 70\% Galaxy mask and point source extractors are used.}}
\label{_pm}
\end{figure}

\begin{figure}[pH]
\centering\includegraphics[width=10.0cm, angle=90.0]{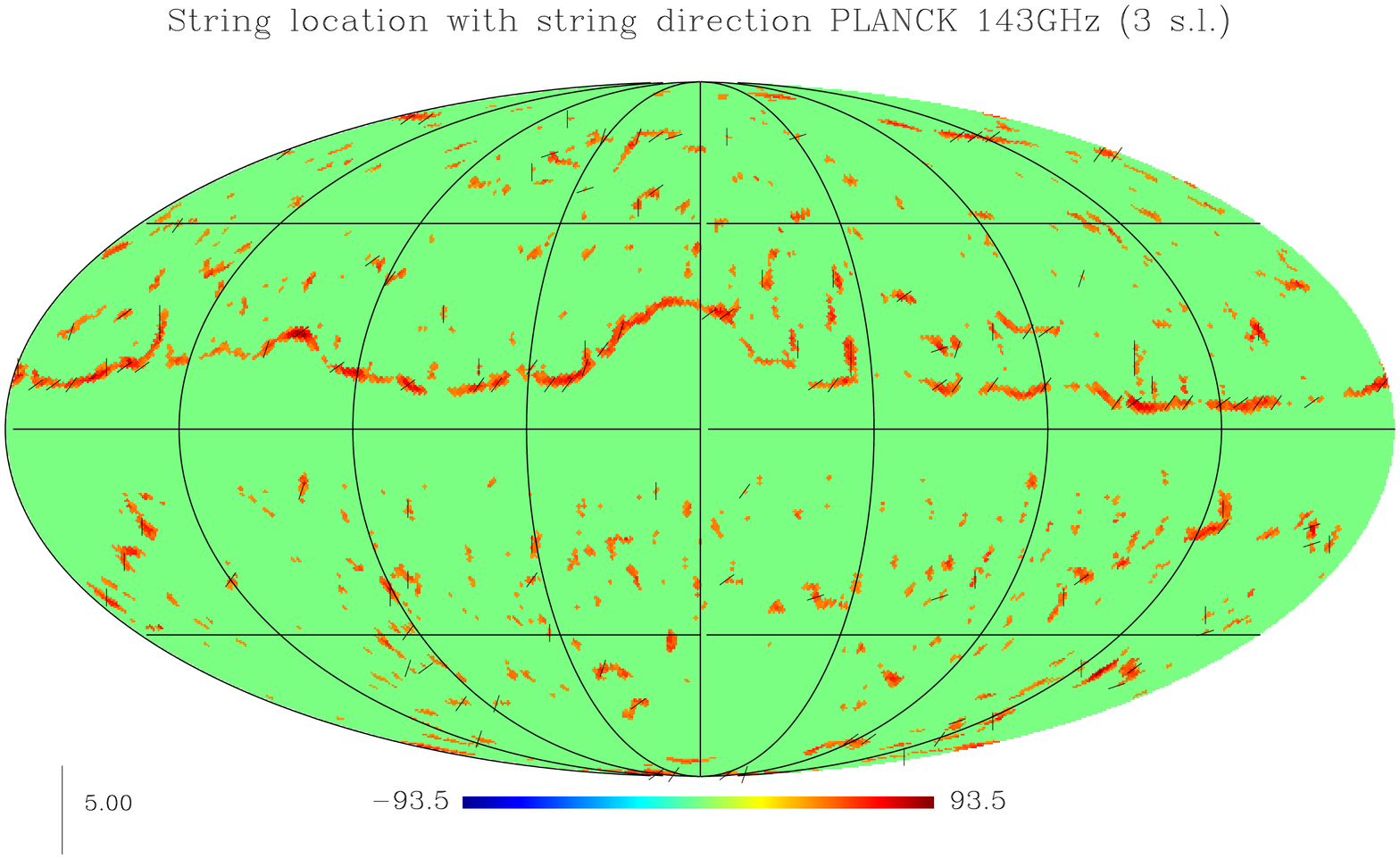}
\caption{\textit{CS candidates (continuous zone with indication of temperature gradients) in Planck data after MHF analysis at the 3$\sigma$ level (see text for details). Units are [$\mu$K]. Radius of MHF convolution is $5^o$. The long continuous traces in the vicinity of the Galactic equator are the remnants of the Galactic filter.}}
\label{_sl}
\end{figure}

We prepared six independent original Planck maps (from 100GHz to 857GHz) cleaning them with recommended Galaxy filters and point source extractors (\cite{11}, see Fig. (\ref{_pm}) as a map example for 143GHz). Than we applied to the cleaned maps the MHF algorithm, convolving them in each pixel with a MHF specified in a circle. As result we find CS candidates (see Fig. (\ref{_sl})). One can see artificial traces along the mask boundary which must be excluded in CS search. The CS number is a function of their tension. The most important conclusion is that we put restrictions on the CS tension.

There are no CS with tension more than $G\mu/c^2 = 7.36 \cdot 10^{-7}$. For tensions in the range $G\mu/c^2 = 6.44 \cdot 10^{-7}$ to $G\mu/c^2 = 7.36 \cdot 10^{-7}$ we have no more than one CS candidate. For the lowest tension limit available by the MHF algorithm we have no more than 5 CS candidates in the whole Universe inside the  last scattering surface. For $G\mu/c^2 \le 4.83 \cdot 10^{-7}$ the MHF method is ineffective because of unverifiable or even wrong CS candidates. Thus the existence of string with tensions $G\mu/c^2 \le 4.83 \cdot 10^{-7}$ is not excluded but it is beyond the Planck data possibilities.

We use simple criteria for a CS candidate. We consider a detection to be positive if we find:
\begin{itemize}
\item a continuous line;
\item at least three correlated vector of temperature gradients.
\end{itemize}

For example, for the filter 143GHz, the $1\sigma$ value corresponds to $\delta T  = 14.8 \mu K$ but in this case we have some wrong candidates which have to be studied by additional optical methods (search for an excess of gravitational lensing events in vicinity of the CS candidates). The $2\sigma$ and $3\sigma$ levels correspond to $29.6 \mu K$ ($G\mu/c^2 = 4.21 \cdot 10^{-7}$, \cite{sup}) and $44.4 \mu K$ ($G\mu/c^2 = 6.32 \cdot 10^{-7}$) respectively.

We have applied this procedure to all the available wave-channels: 
100GHz, 143GHz, 217GHz, 353GHz, 545GHz, 857GHz. We have applied all the available point source masks for each channel and Galaxy masks (to extract the Galaxy radiadions) for recommended \cite{11} sky coverage of 70\%, 80\%, 90\%, 95\%, 97\%, 99\% (Table (\ref{_tab1})). 

Of course, in this procedure we can miss some candidates lying in the equatorial Galactic region.

\begin{table} 
\begin{center}
\begin{tabular}{|c|c|c|} 
\hline 
CS candidate number        & CS tension     &  Sky coverage \\ 
\hline 
3  & 5.52 & 97 \\
2  & 5.66 & 99 \\ 
2  & 6.15 & 90 \\
2  & 6.32 & 70 \\
1  & 7.07 & 99 \\
1  & 7.36 & 97 \\
\hline
\end{tabular}\caption{The result of CS candidates search by the MHF algorithm applied to Planck CMB data is shown for filter 143GHz. 
The first column gives the number of CS candidates with given tension $G\mu/c^2$ (second column, in $10^{7}$) for different sky coverage (third column, in percents). The sky coverage characterizes the type of Galactic mask.}\label{_tab1}
\end{center}
\end{table}
We used all available filters to compare the positions of candidates and filter out those which are not present at all frequencies, as the appearance of a real CS shall not depend on the observation frequency. 

It should be emphasized we found CS candidates in two independend data sets: WMAP and Planck.

\section{Restrictions on the CS numbers}

In the previous Sections it was obtained the general restriction on CS tension and their number. 

Let us now analyse the different CS types. 

The tensions of solitary CS candidates (Table (\ref{_tab1})) can be compared with upper bounds of CS tensions found by Planck team \cite{11} based on CS network simulations (Table (\ref{_tab2})). 
\begin{table} 
\begin{center}
\begin{tabular}{|c|c|c|}
\hline 
CS network    & Data &  CS tension \\ 
\hline 
Nambu-Goto model  & Planck + WP  & 1.5 \\ 
\hline 
Abelian-Higgs field theory model & Planck + WP & 3.2 \\ 
\hline 
Abelian-Higgs mimic model & Planck + WP         &  3.6  \\ 
\hline
Semilocal CS model  & Planck + WP         &  11.0 \\
\hline
Global texture model  & Planck + WP         &  10.6 \\
\hline
\end{tabular}\caption{The upper bounds on CS and textures tension $G\mu/c^2$ (third column, in $10^{7}$) for different types of CS network and texture simulations (first column) using combined CMB data from Planck and WMAP polarization (\cite{11}).}\label{_tab2}
\end{center}
\end{table}

Let us suppose, without loss of generality, a homogeneous distribution of CS in the network. In fact this assumption did not contradict with observation and there are a lot of types of networks in theoretical approaches. Therefore, for solitary CS the restriction on tension becomes:
$$
\biggl( G \mu/c^2 \biggr)|_{\mathit{solitary}} \sqrt{N} = \biggl( G \mu/c^2 \biggr)|_{\mathit{network}},
$$
where $N$ is CS number. 

To estimate the contribution of the energy of the CS network on the total energy of the Universe it is usually used the unequal time correlator (UETC) of the CS stress energy tensor \cite{ue}. The CS tension is usually quantified in terms of the dimensionless ratio $G\mu /c^2$. This ratio is not a CS tension of one string. This is only normalization of power spectrum produced by CS network to be consistent with CMB data \cite{ri}. In other words this value gives us the upper limit to estimate the fraction of energy in string with respect to the total energy of the Universe. 

Planck (and WMAP) cannot mark out single strings. They are dealing with the network in a whole. Values from Table (\ref{_tab2}) (see \cite{11} for details) give the inequalities for each CS types:
\begin{eqnarray}
\biggl( G \mu/c^2 \biggr)|_{\mathit{solitary}} \sqrt{N}<a,
\label{_in}
\end{eqnarray}  
where $a$ is the corresponding upper limit from Table (\ref{_tab2}).

The MHF allow us to try to find individual, single strings. Therefore we can compare the tension from network simulations with tension of individual CS. 

\begin{figure}[pH]
\centering\includegraphics[width=10.0cm]{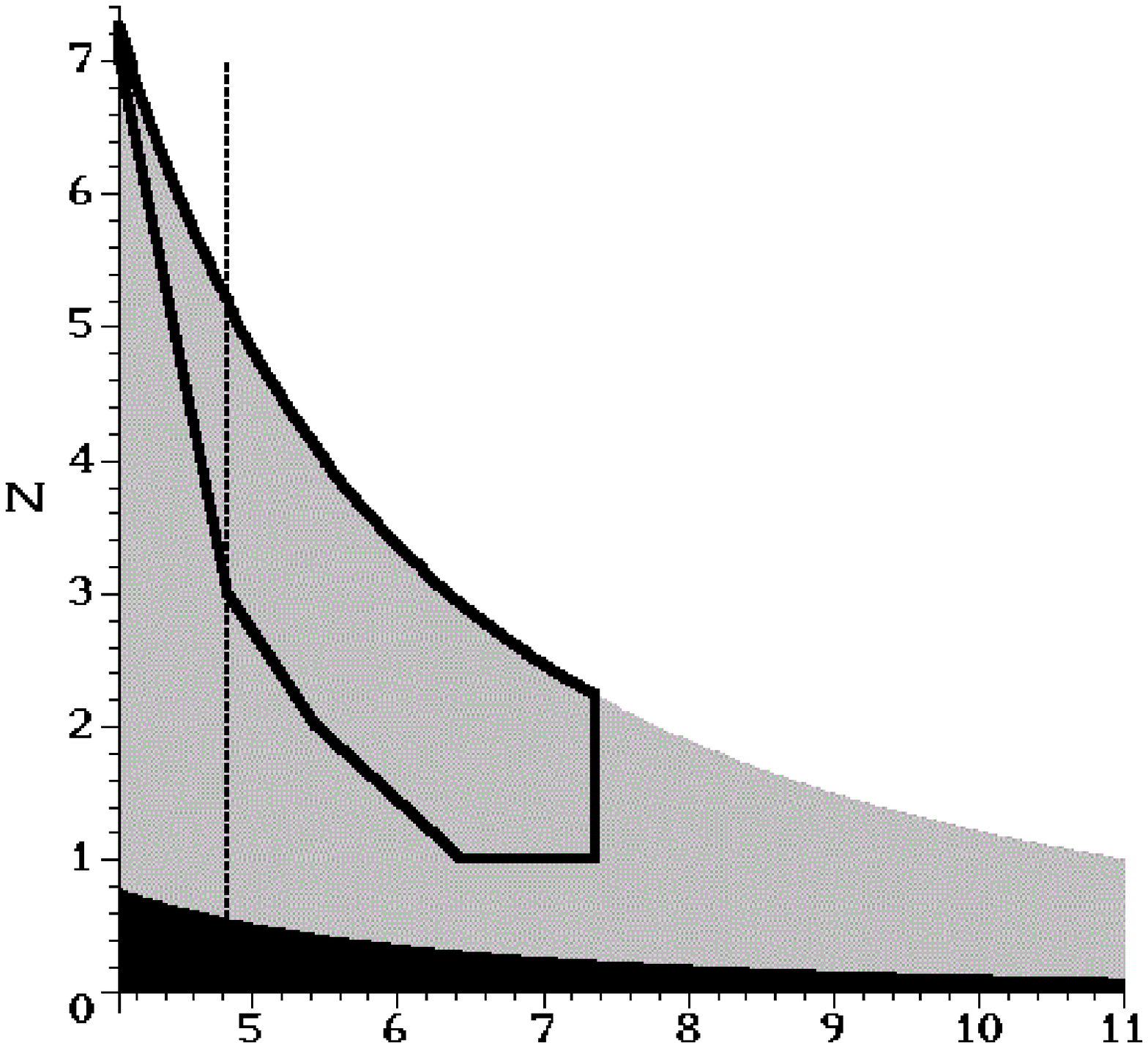}
\caption{\textit{The permitted region for CS to exist (see text for details). The string number $N$ vs string tension (in $G\mu/c^2 \cdot 10^{7}$). }}
\label{_11}
\end{figure}

In the Fig. (\ref{_11}) we put together our CS candidates and regions from (\ref{_in}). 

The vertical axis is CS number. The horizontal axis is CS tension $G \mu/c^2$ in units $10^{7}$. The gray region is that permitted for semilocal CS and textures model. The black region is that permitted for Nambu-Goto, Abelian-Higgs field theory, and Abelian-Higgs mimic CS models. The areas were obtained from the inequality (\ref{_in}).

The closed black contour is from Planck CMB data analysis by MHF. One can see the only semilocal CS (and textures) models are consistent with estimation of solitary CS number. Therefore, the main result is that in MHF analysis there are no CS candidates of Nambu-Goto, Abelian-Higgs field theory, and Abelian-Higgs mimic models. The region at the left side of the vertical dash line is where the MHF method is ineffective, as we already mentioned. Thus the existence of semilocal CS with tensions $G\mu/c^2 \le 4.83 \cdot 10^{-7}$ it is not forbidden but it is beyond the Planck data scope.

\section{Conclusions}

It was elaborated, simulated and tested new algorithm to CS detection in CMB maps (WMAP, Planck). This algorithm is based on convolution procedure of original observational radio data with modified Haar functions (MHF) and is able to achieve the resolution for CS deficit angles of the order of 1 arcsec. The main result that there are no CS with tension more than $G\mu/c^2 = 7.36 \cdot 10^{-7}$.

Secondly, it was found a list of preliminary CS candidates in Planck data. The same CS candidates have been found in the WMAP 9-yr data. Independence of Planck and WMAP data sets serves as an additional argument to consider those CS candidates as very promising. However the final proof should be given by optical deep surveys by observation of gravitational lensing chains (this is the aim of future intense work).

Finally, our MHF algorithm with the results in \cite{11} made it possible to clarify the preferred CS types. The most preferable types of CS are semilocal ones, described by the model  with complex scalar dublet \cite{8}. If its imaginary part is equal to 0, the semilocal CS becomes the Abelian-Higgs CS. The main difference between these two types of CS is that the semilocal CS can have ends (monopoles) and can be unstable under certain conditions. The topological (''ordinary'') CS have no ends. Formally they break on the surface of last scattering. It means that if our CS candidates are topological defects, then they have to be very far from the observer, up to z=7, because their length is much less than $100^{0}$, \cite{7}. In this case we have no possibility to observe their effects in the optical data by looking through gravitational lensing events, and we will never confirm our candidates by independent optical observation. But the situation substantially changes if we are dealing with semilocal CS. They can be closer to us, being not very long. Therefore our strategy  is now to find suitable optical fields to search for the chains of gravitational lenses, produced by candidates to semilocal CS. The structure of CS candidates found by the MHF method confirms the view of semilocal CS as a collection of segments. It is also necessary to mention that the comparisons with the data for the search of non-Gaussian signals (\cite{11}) shows that the presence of several (up to 3) Nambu-Goto CS is also possible but the semilocal CS remain the most favoured.

\vskip3mm
We are very grateful to Prof. M. Capaccioli for discussion. The support by the grant "Messaggeri della conoscescenza" of the Italian Ministry of University and Research is acknowledged.

\section*{Appendix I. Modified Haar functions with cyclic shift}
According general theorems for Euclidean spaces, in space $L_{2}$ there are complete orthogonal systems of functions. 

Any function $g(x)$ belonging to the space $L_2$ can be represented as a sum of Fourier series on a system of functions $f_i(x)$:

\[
g(x) = \sum_{i=1}^{\infty} c_i f_i(x).
\] 

Here $c_i$ are Fourier coefficients of the function $g(x)$ on the system $f_i(x)$:

\[
c_i = \frac{1}{\left\| f_i \right\|^2} \int{ g(x)f_i(x)d\mu }, 
\]

where

\[
\left\| f_i \right\|^2 = \int{f_i(x)^2 d\mu}. 
\]

By Haar \cite{kolm} it has been introduced a set of functions, complete and orthonormal on the space $[0,1]$. This system has been defined by function: 

\[
\phi_0 = 1
\]

and by set of functions (''series'' of functions):

\begin{eqnarray*}
\phi_{0 \,1} \\
\phi_{1 \,1}, \phi_{1 \,2} \\
\phi_{2 \,1}, \phi_{2 \,2}, \phi_{2 \,3}, \phi_{2 \,4} \\
\ldots \\
\phi_{n \,1}, \phi_{n \,2}, \phi_{n \,3}, \ldots, \phi_{n \,2^n}.
\end{eqnarray*}

The ''series'' number $n$ contains $2^n$ functions. 

\begin{eqnarray*}
{\phi _{0 \,1}} = 
\begin{cases}
1,  0 < x < 1/2, \\ 
-1, 1/2 < x < 1, 
\end{cases}
\end{eqnarray*}

\begin{eqnarray*}
{\phi _{1 \,1}} = 
\begin{cases}
\sqrt{2}, 0 < x < 1/4, \\ 
-\sqrt{2}, 1/4 < x < 1/2, \\
 0,        1/2 < x < 1,
\end{cases}
\end{eqnarray*}

\begin{eqnarray*}
{\phi _{1 \,2}} = 
\begin{cases}
0,        0 < x < 1/2, \\ 
\sqrt{2},  1/2 < x < 3/4, \\
 -\sqrt{2}, 3/4 < x < 1.
 \end{cases}
\end{eqnarray*}

In points of discontinuity the Haar funcions $\phi_{n, \,i}$ can be defined arbitrarily. In the common case:
\[
\phi_0 = 1
\]
\begin{eqnarray*}
{\phi _{n \,i}} = 
\begin{cases}
 2^{\frac{n}{2}},  {\displaystyle \frac{i-1}{2^n}} < x < {\displaystyle\frac{i-1}{2^n}+\frac{1}{2^{n+1}}}, \\ 
-2^{\frac{n}{2}}, {\displaystyle\frac{i-1}{2^n}+\frac{1}{2^{n+1}}} < x < {\displaystyle\frac{i}{2^n}}, \\
               0,       x \notin {\displaystyle\biggl[ \frac{i-1}{2^n};\frac{i}{2^n} \biggr]},
\end{cases}                           
\label{sys}
\end{eqnarray*}
\[
n=0,1, \ldots; i=1,2, \ldots,2^n.
\]

Let us introduce the modified Haar functions $\{\psi_{n \, i} \}$ with cyclic shift \cite{9}. For simplicity we will take them on the compact space $[0,1]$ with real cyclic shift $a \in [0, 1/2]$. Depending on the parameter $a$ the functions $\{\psi_{n \, i} \}$ can be divided into similar four groups for $0< a < 1 - i/2^n$, $1 - i/2^n < a < 1 - i/2^n + 1/2^{n+1}$, $1 - i/2^n + 1/2^{n+1}< a < 1 - i/2^n + 1/2^{n}$, and $1 - (i-1)/2^n < a < 1/2$. 

The subscripts ''0'' and ''1'' refer to the radial and angular variables of a CS, respectively.

For the first case we have:
\begin{eqnarray*}
{\psi _{n \,i}^{(a)}} =
\begin{cases}
2^{\frac{n}{2}},  {\displaystyle \frac{i-1}{2^n}+a} < x < {\displaystyle \frac{i-1}{2^n}+a+\frac{1}{2^{n+1}}}, \\
-2^{\frac{n}{2}}, {\displaystyle\frac{i-1}{2^n}+a+\frac{1}{2^{n+1}}} < x < {\displaystyle\frac{i}{2^n}+a}, \\
0,        x \notin {\displaystyle\biggl[ \frac{i-1}{2^n}+a;\frac{i}{2^n}+a \biggr]}.
\end{cases}                  
\label{sys1}
\end{eqnarray*}

If $1 - i/2^n < a < 1 - i/2^n + 1/2^{n+1}$, then
\begin{eqnarray*}
{\psi _{n \,i}^{(b)}} = 
\begin{cases}
 2^{\frac{n}{2}},  {\displaystyle\frac{i-1}{2^n}+a} < x < {\displaystyle\frac{i-1}{2^n}+a+\frac{1}{2^{n+1}}}, \\
-2^{\frac{n}{2}},  {\displaystyle\frac{i-1}{2^n}+a+\frac{1}{2^{n+1}}} < x < {\displaystyle 1} \,\, \bigcup \,\, {\displaystyle 0} < x < {\displaystyle \frac{i}{2^n}+a-1}, \\
               0,  x \in {\displaystyle \biggl[ \frac{i}{2^n}+a-1;\frac{i-1}{2^n}+a \biggr]}.
\end{cases}
\label{sys2}
\end{eqnarray*}

If $1 - i/2^n + 1/2^{n+1}< a < 1 - i/2^n + 1/2^{n}$, then
\begin{eqnarray*}
{\psi _{n \,i}^{(c)}} = 
\begin{cases}
2^{\frac{n}{2}}, {\displaystyle \frac{i-1}{2^n}+a} < x < {\displaystyle 1} \,\, \bigcup \,\, {\displaystyle 0}<x< {\displaystyle \frac{i-1}{2^n}+a + \frac{1}{2^{n+1}} -1}, \\ 
-2^{\frac{n}{2}}, {\displaystyle \frac{i-1}{2^n}+a+\frac{1}{2^{n+1}}-1} < x < {\displaystyle\frac{i}{2^n}+a-1}, \\
0,        x \in {\displaystyle \biggl[ \frac{i}{2^n}+a-1;\frac{i-1}{2^n}+a \biggr]}.
\end{cases}
\label{sys3}
\end{eqnarray*}

If $1 - (i-1)/2^n < a < 1/2$, then
\begin{eqnarray*}
{\psi _{n \,i}^{(d)}} = 
\begin{cases}
2^{\frac{n}{2}},  {\displaystyle \frac{i-1}{2^n}+a-1} < x < {\displaystyle\frac{i-1}{2^n}+a-1+\frac{1}{2^{n+1}}}, \\ 
-2^{\frac{n}{2}}, {\displaystyle\frac{i-1}{2^n}+a-1+\frac{1}{2^{n+1}}} < x < {\displaystyle\frac{i}{2^n}+a-1}, \\
                          0,       x \notin {\displaystyle\biggl[ \frac{i-1}{2^n}+a-1;\frac{i}{2^n}+a-1 \biggr].}
\end{cases}
\label{sys4}
\end{eqnarray*}

For fixed $a$ the set of modified Haar functions is complete and orthonormal, as well as the system of classical Haar functions \cite{9}. Therefore it can be correctly used in searching of signals.

\section*{Appendix II. Simulations of false string detection}

\begin{figure}[pH]
\centering\includegraphics[width=10.0cm, angle=90.0]{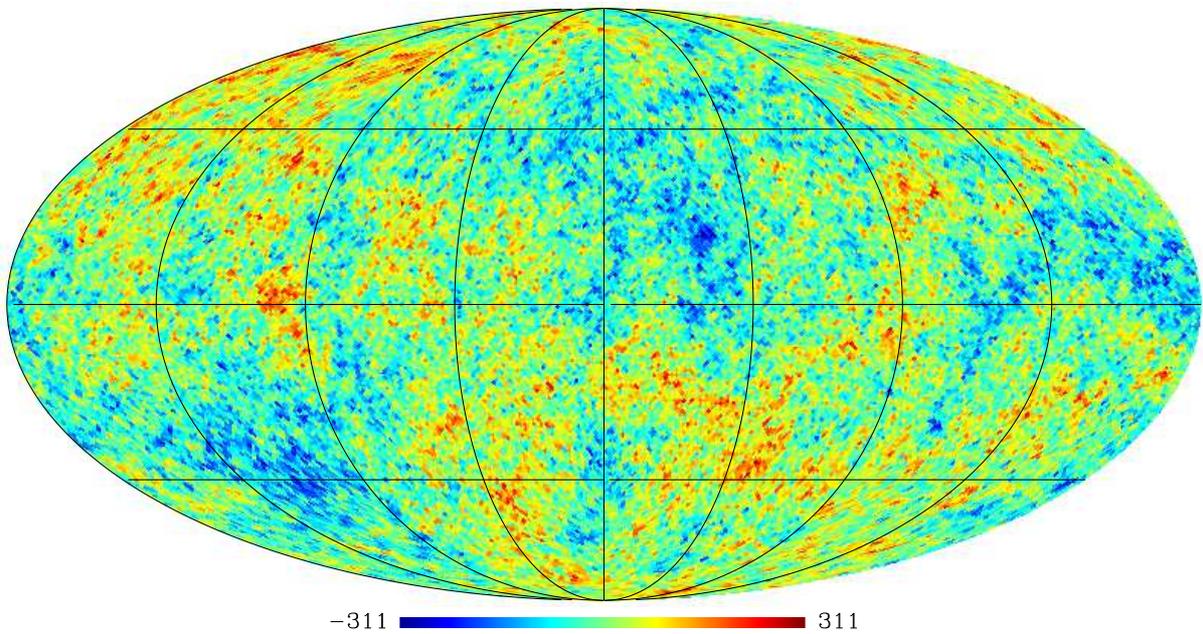}
\caption{\textit{Simulated CMB map. Example (1). Units are [$\mu$K].}}
\label{_01}
\end{figure}

\begin{figure}[pH]
\centering\includegraphics[width=10.0cm, angle=90.0]{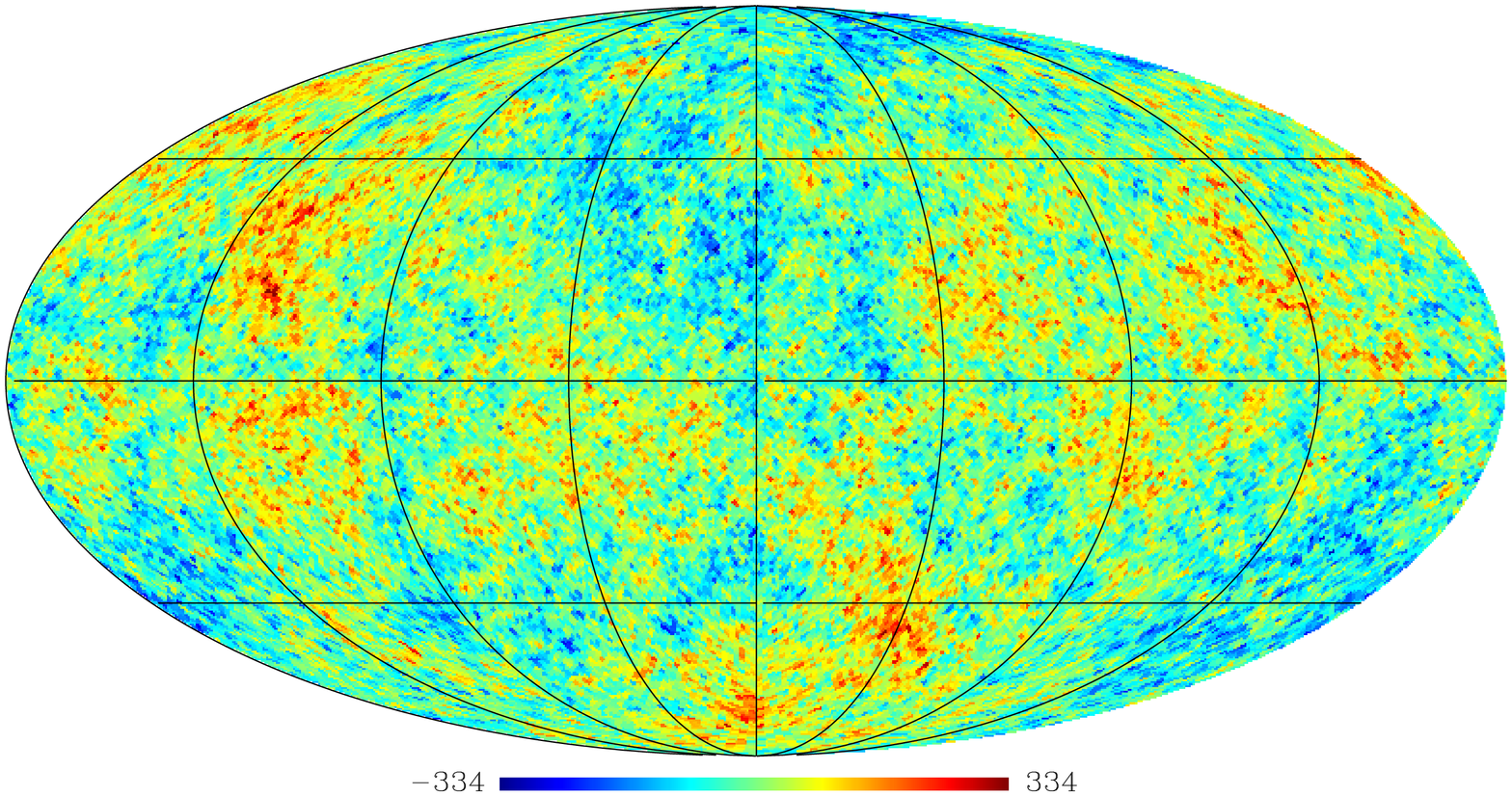}
\caption{\textit{Simulated CMB map. Example (2). Units are [$\mu$K].}}
\label{_02}
\end{figure}

\begin{figure}[pH]
\centering\includegraphics[width=10.0cm, angle=90.0]{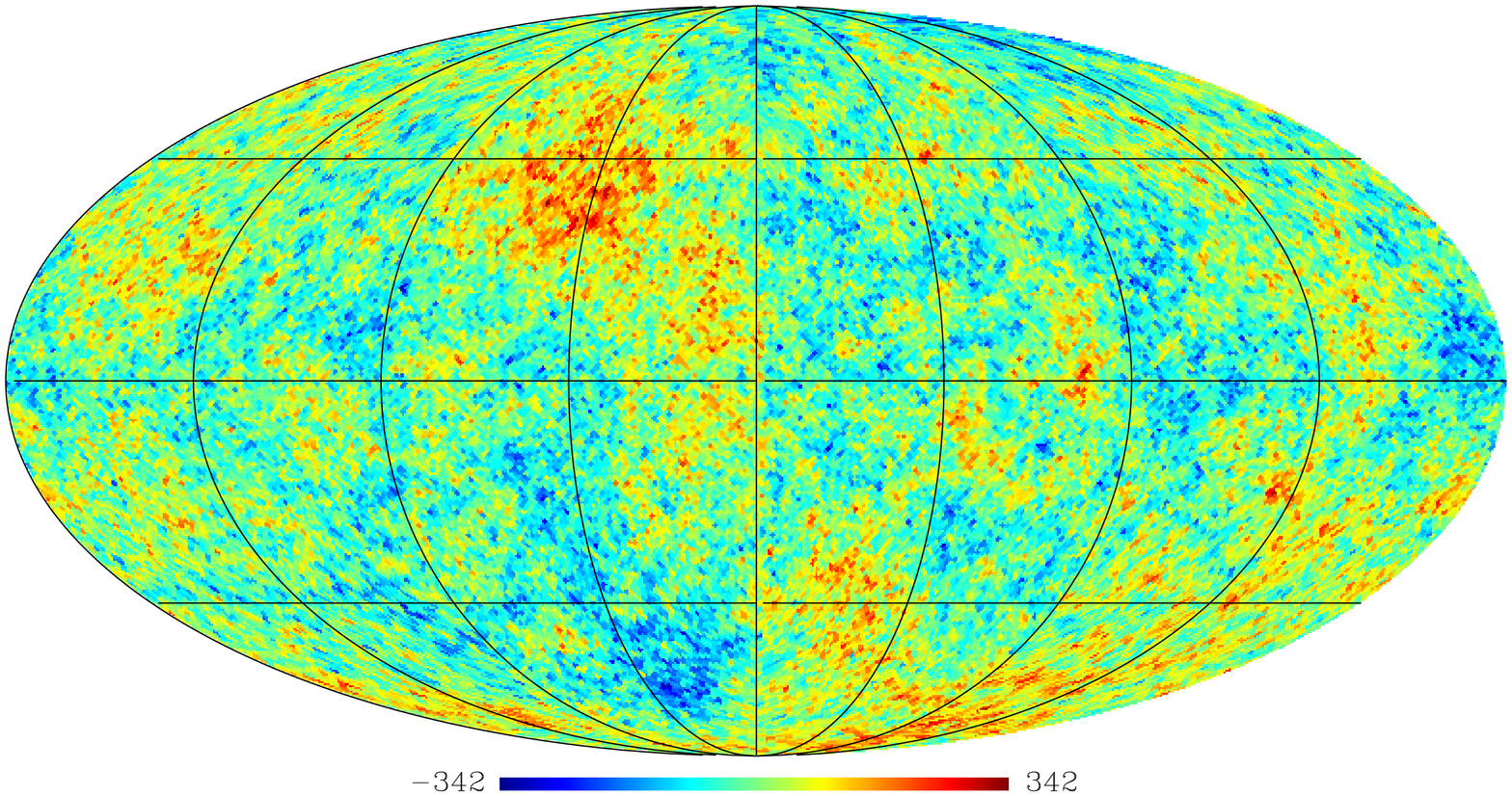}
\caption{\textit{Simulated CMB map. Example (3). Units are [$\mu$K].}}
\label{_03}
\end{figure}

\begin{figure}[pH]
\centering\includegraphics[width=10.0cm, angle=90.0]{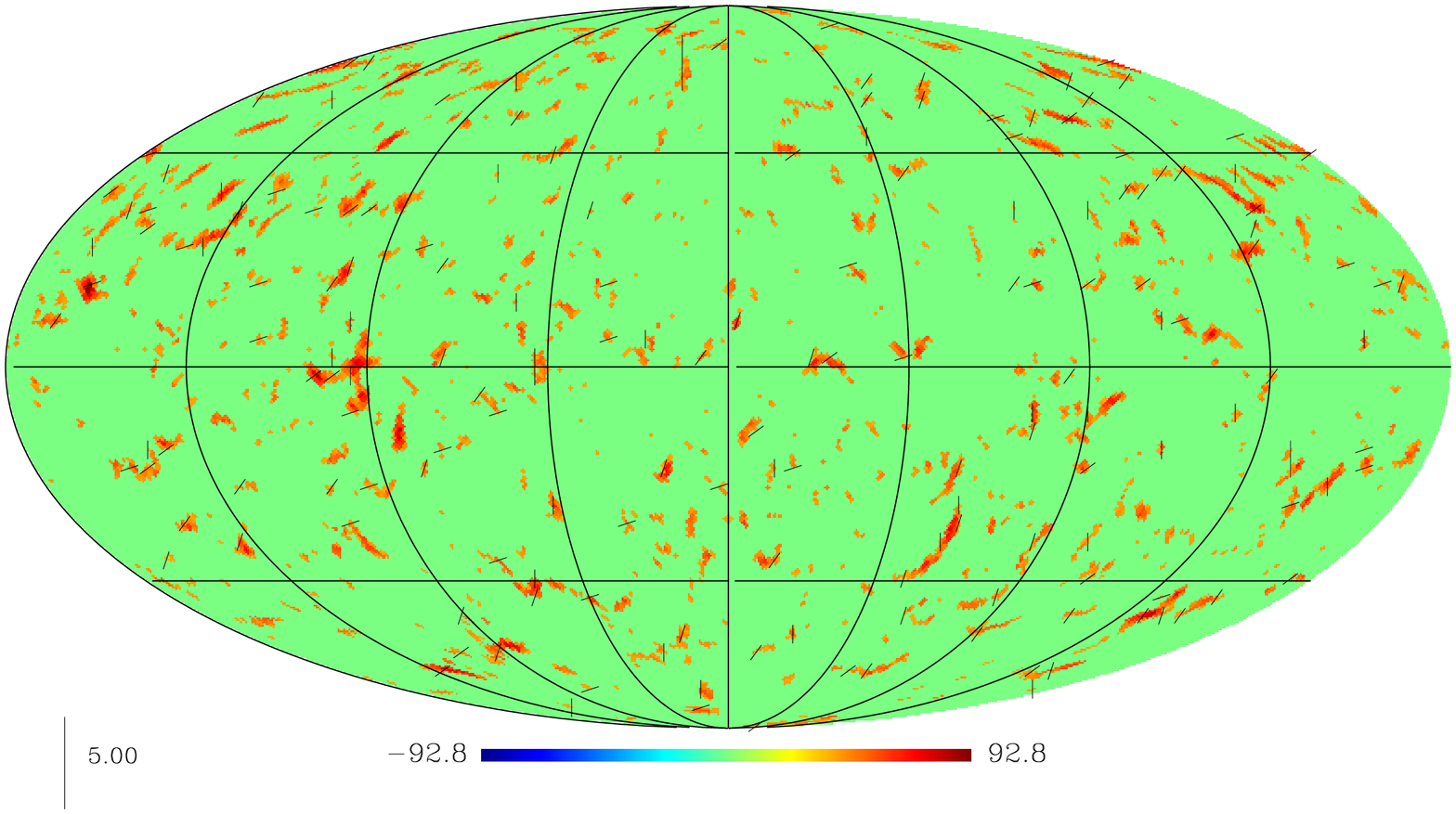}
\caption{\textit{False CS candidates for simulated CMB map on the Figure. (\ref{_01}). There is one false CS candidate under two necessary conditions for the existence of strings (Sec. \ref{cs_}). Units are [$\mu$K].}}
\label{_1}
\end{figure}

\begin{figure}[pH]
\centering\includegraphics[width=10.0cm, angle=90.0]{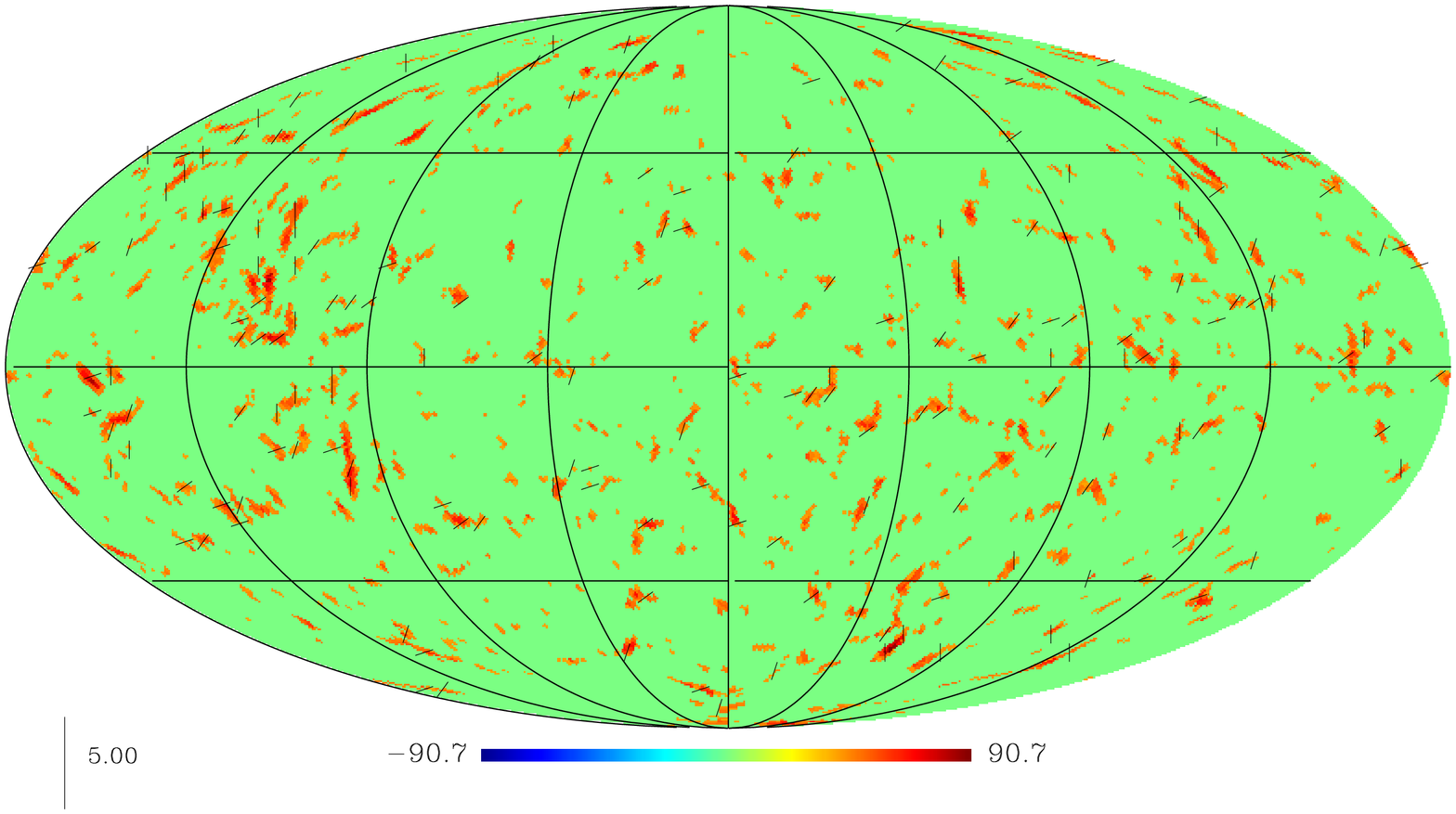}
\caption{\textit{False CS candidates for simulated CMB map on the Figure. (\ref{_02}). There is no false CS candidate under two necessary conditions for the existence of strings (Sec. \ref{cs_}). Units are [$\mu$K].}}
\label{_2}
\end{figure}

\begin{figure}[pH]
\centering\includegraphics[width=10.0cm, angle=90.0]{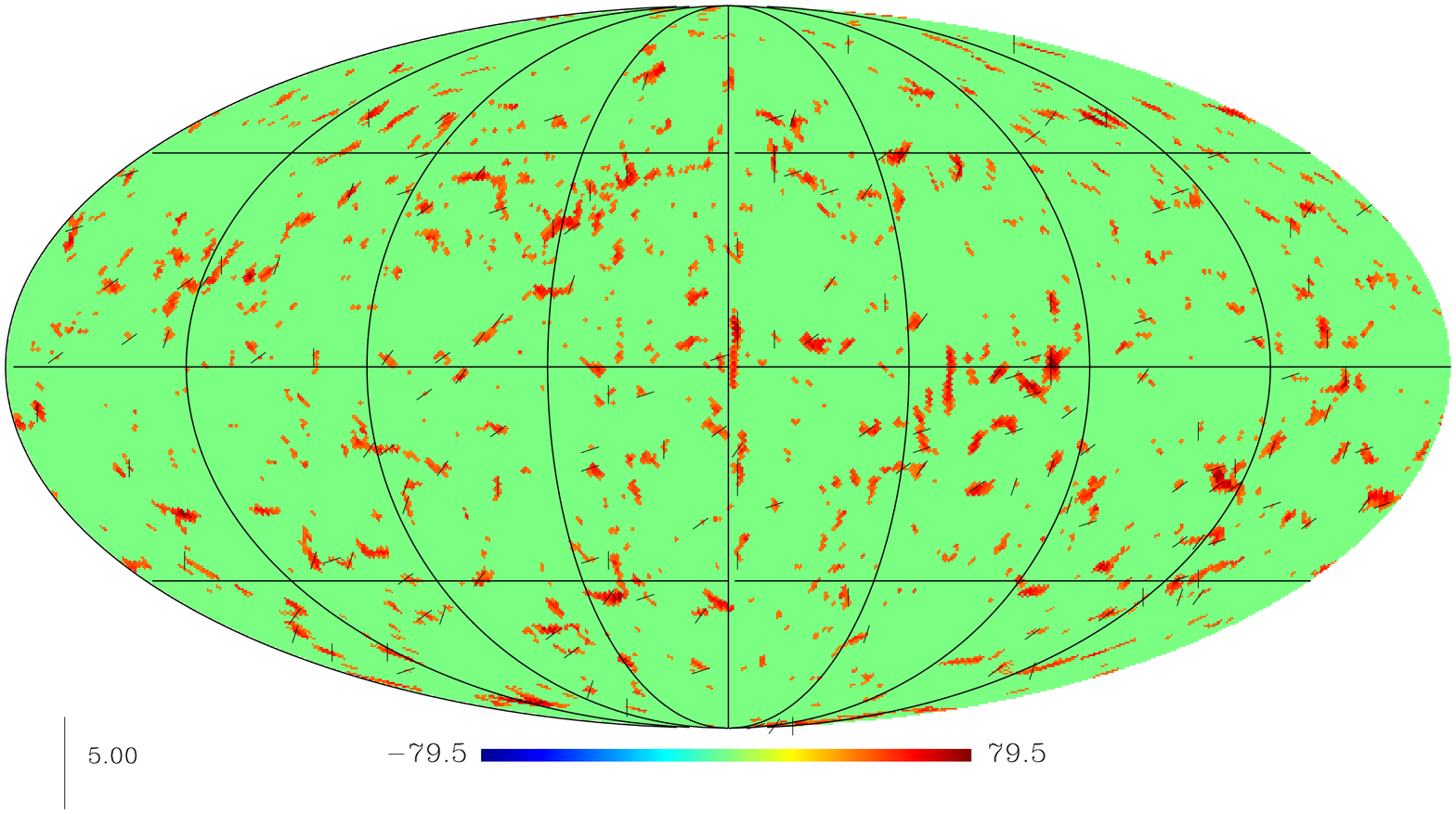}
\caption{\textit{False CS candidates for simulated CMB map on the Figure. (\ref{_03}). There is one false CS candidate under two necessary conditions for the existence of strings (Sec. \ref{cs_}). Units are [$\mu$K].}}
\label{_3}
\end{figure}

In order to study the efficiency of the MHF algorithm we generate 300 maps of sky simulating the CMB structure without any string. In Figs. (\ref{_01}) - (\ref{_03}) we show examples of those simulated maps, in Figs. (\ref{_1}) - (\ref{_3}) we show examples of the result of the MHF algorithm.


Let us estimate the average number of false CS candidates (i.e. ''artifacts'') satisfying nonetheless, the two necessary conditions for the existence of strings (Sec. \ref{cs_}): 
\begin{itemize}
\item a continuous line;
\item at least three correlated vector of temperature gradients.
\end{itemize} 

The analysis of 50 simulated maps gives the expected number of false CS candidates in the whole sky as 2.3 with low despersion. If we use the Galactic filter, 70\% and 90\%, we obtain 0.69 and 0.23 averages numbers of false CS candidates respectively. Analysis of original observational WMAP and Planck data shows the presence of 1 up to 5 CS candidates when using 70\% of the Galactic filter (recommended by \cite{11}). Significance level is 3$\sigma$.

If in the data (using 70\% Galactic filter) is found only one CS candidate, then the probability that this is a false candidate is 26\%. If there are two CS candidates in the data (using the same 70\% Galactic filter), the statistics on false candidates not explain this excess.



\begin{thebibliography}{99}
\bibitem{1}
T.W.B.Kibble, Topology of cosmic domains and strings, J.Phys.A:Math.Gen. (1976) 9.
\bibitem{2}
Ya.B.Zeldovich, Cosmological fluctuations produced near a singularity, MNRAS 192 (1980) 663.
\bibitem{3}
A.Vilenkin, Gravitational field of vacuum domain walls and strings, Phys.Rev.D 23 4 (1981) 852.
\bibitem{31}
J.Urrestilla, et al., Cosmic microwave anisotropies from BPS semilocal strings, JCAP 0807 (2008) 010.
\bibitem{ve}
O. S. Sazhina, Probabilistic Estimates of the Number of Cosmic Strings, (in russian) JETP 116 1 (2013) 71�79. 
\bibitem{vil}
A.Vilenkin, Cosmic strings as gravitational lenses, Ap. J.  L51 (1984) 282.
\bibitem{vil2}
A.Vilenkin, Looking for cosmic strings, Nature 322 (1986) 613.
\bibitem{4}
M.V.Sazhin, et al., CSL-1: chance projection effect or serendipitous discovery of a gravitational lens induced by a cosmic string? MNRAS 343 2 (2003) 353.
\bibitem{5}
M.V.Sazhin, M.Capaccioli, G.Longo, M.Paolillo, O.S.Khovanskaya, Further spectroscopic observations of the CSL-1 object, Astrophys.J. 636 (2005) L5-L8.
\bibitem{6}
M.V.Sazhin, O.S.Khovanskaya, et al., Gravitational lensing by cosmic strings: What we learn from the CSL-1 case, MNRAS 376 (2007) 1731.
\bibitem{71}
N. Kaiser and A. Stebbins, Microwave Anisotropy due to Cosmic Strings, Nature 310 (1984) 391-393.
\bibitem{72}
A. Stebbins, Cosmic strings and the microwave sky. 1: Anisotropy from moving strings, Ap. J. 327 (1988) 584.
\bibitem{7}
O.S.Sazhina, M.V.Sazhin, V.N.Sementsov, Anisotropy of CMBR induced by a straight moving cosmic string, (in russian), JETP 133 5 (2008) 1005.
\bibitem{ue}
Ue-Li Pen, Uro's Seljak, and Neil Turok, Power Spectra in Global Defect theories of Cosmic Structure Formation, astr-ph 9704165v2.
\bibitem{ri}
Richard Battye, and Adam Moss Updated constraints on the cosmic string tension, astro-ph1005.0479v2.

\bibitem{8}
A.Vilenkin and E.P.Shellard, Cosmic strings and other topological defects, Cambridge Univ.Press, UK 1994.
\bibitem{91}
O.S.Sazhina, V. N. Sementsov, and N. T. Ashimbaeva, Cosmic String Detection in Radio Surveys, Astron. Rep. 58 1 (2014) 16 ½29.
\bibitem{9}
O.S.Sazhina, Search for cosmic strings by modified Haar functions with cyclic shift, (in russian), Vestnik MSU 6 (2011) 588.
\bibitem{sup}
E.J. Copeland and T.W.B. Kibble, Cosmic Strings and Superstrings, arXiv:0911.1345v3 [hep-th] 26 Nov 2009.
\bibitem{11}
Planck Collaboration: P. A. R. Ade at al., Planck 2013 results. XXV. Searches for cosmic strings and other topological defects, Astronomy and Astrophysics manuscript no. Defects March 22, 2013.
\bibitem{kolm}
Elements of the Theory of Functions and Functional Analysis 
by A. N. Kolmogorov and S. V. Fomin, Graylock Press, Rochester, N.Y. 1957.

\bibitem{q1}
Riccio G., D'Angelo G., Sazhin M.V., Sazhina O. S., Longo G. and Capaccioli M. Simulations of cosmic strings signatures in the CMB. FINAL WORKSHOP OF GRID PROJECTS, PON RICERCA 2000-2006, AVVISO 1575 (2009) and references therein.
\bibitem{q2}
http://www.thphys.uni-heidelberg.de/~robbers/cmbeasy/
\bibitem{q3}
http://www.cfa.harvard.edu/
\end{thebibliography}
\end{document}